\documentclass[fleqn,12pt,twoside]{article}
\usepackage{espcrc1}

\usepackage{graphicx}
\usepackage[figuresright]{rotating}

\newcommand{\AmS}{{\protect\the\textfont2
  A\kern-.1667em\lower.5ex\hbox{M}\kern-.125emS}}

\title{Concluding remarks on QCD-N'02 Workshop in Ferrara}

\author{G. van der Steenhoven\address[NIKHEF]{Nationaal Instituut voor
	Kernfysica en Hoge-Energiefysica (NIKHEF),\\ 
        P.O. Box 41882, 1009 DB Amsterdam, The Netherlands}}%
       
\begin{document}

\maketitle

\begin{abstract}
A summary is given of some important developments in QCD
studies of the nucleon as presented at this workshop. Based on these
developments some expectations for the short- and long-term future of the
field are sketched. Taken together, the summary of the workshop and the
future perspectives result in a {\sl Road Map} for experimental studies
of the QCD structure of the nucleon. The Road Map includes as a long-term
goal the construction of new lepton-hadron scattering facilities both
in Europe and the United States.
\end{abstract}

\section{INTRODUCTION}

The workshop on {\sl The QCD Structure of the Nucleon} in Ferrara was
organized in order to address the status and the future of the field, with 
the aim of arriving at a {\sl Road Map} describing where and how to go in the 
next decade. The goal of the conference was ambitious, but well motivated.
In recent years studies of QCD aspects of nucleon structure are faced with 
a number of striking developments, which are holding promise for the future:
\begin{itemize}
\item{The renewed interest in sofar unmeasured distribution and fragmentation 
functions, in particular the transversity distribution $h_1(x)$, which enable 
us to obtain experimental information on new aspects of nucleon structure such
as the tensor charge of the nucleon.}
\item{The introduction of Generalized Parton Distributions (GPDs), which 
contain information on dynamic correlations between partons in the nucleon. 
The GPDs provide a unified framework that can be used for the description
of a wide range of observables, and which encompass the well-known structure 
functions and nucleon form factors as limiting cases. This framework
can also be used to demonstrate how deeply virtual Compton scattering
(DVCS) enables us to determine the total angular momentum ($J$) carried
by the partonic constituents of the nucleon.}
\item{First observations of single-spin asymmetries in deep inelastic 
lepton scattering, demonstrating that -- at least in principle -- both
the transversity distribution and the generalized parton distributions
can be accessed experimentally.}
\end{itemize}
While these developments issue new avenues in our field, some older
subjects are going through a phase of rapid changes as well. Examples
of such developments include lattice gauge theory, which is now
being able to make a closer link to experimental observables, and 
the observation of new evidence in favour of Colour Transparency, 
a long-standing QCD-based prediction.

In these concluding remarks I will summarize these developments using
the  material presented at the workshop. This is the subject of section 2.
In section 3 some -- more subjective -- impressions are given on the
prospects of our field in the future. Which issues will most likely
be resolved, and where should be the emphasis of our research efforts?
The first preliminary data on the new observables mentioned above imply that 
new lepton-scattering facilities of high luminosity are needed in order
to be able to carry out measurements of reasonable precision. A road-map
that might lead to the realization of such facilities is sketched in
the concluding section of this paper.

\section{SUMMARY OF THE WORKSHOP}

In the opening talk of the workshop P. Hoyer~\cite{Hoye02} described the
fundamental problem addressed in our field as {\sl "finding the
structure of relativistic QCD bound states in a gluon condensate with
spontaneously broken chiral symmetry"}. The importance of this problem
was illustrated by mentioning that 98\% of the mass of the proton 
originates from QCD binding effects, while only 2\% is due to the
Higgs mechanism. Hoyer asked whether a solution of this problem was within
reach, and in answering this question he listed most of the 
aforementioned  subjects using them as a sign for a promising future. 
These subjects are summarized below in separate subsections. 

It is noted that necessarily only a subset of all material
presented at the workshop is represented, and for illustrative figures
the reader is referred to the original papers which are also contained
in this volume.

\subsection{The transversity distribution}

Apart from the structure functions $F_{1,2}(x)$ and $g_{1,2}(x)$, there
is a third leading-twist structure function $h_1(x)$ that is known
as the transversity distribution. It is of great interest
to measure $h_1(x)$, for which no data exist, since the transverse
spin structure of the nucleon is expected to be considerably different
from the longitudinal spin structure. Both chiral-soliton (instanton)
models~\cite{Chiral} and lattice gauge calculations~\cite{Lattice}
predict that the tensor charge $\delta\Sigma_q$ is considerably larger 
than the longitudinal quark spin contribution $\Delta\Sigma_q$. This
is caused by the absence of gluon-splitting in the transverse
case which results in a predicted relatively weak $Q^2$ dependence of
$h_1(x)$.
 
Inclusive deep-inelastic scattering cannot be used to measure $h_1(x)$ 
as it is a chirally-odd quantity. In semi-inclusive DIS information
on $h_1(x)$ can be obtained because it appears in the cross section
expression in combination with a chirally-odd fragmentation
function~\cite{MT}. First evidence of a non-zero transversity distribution
has been reported by HERMES~\cite{H-ssa}. In this experiment the single
target-spin asymmetry for leptoproduction of pions was measured
on a longitudinally polarized hydrogen target. The data show a
small semi-inclusive asymmetry. It can be explained from the small 
transverse polarization component of the virtual
photon combined with reasonable non-zero values for $h_1(x)$ and
the corresponding chirally-odd fragmentation function (Collins 
effect~\cite{Coll93}). 

During the workshop an interesting discussion emerged concerning the
interpretation of the single-spin asymmetries observed by HERMES. It
was argued by Hwang~\cite{Hwang} that the data could also be explained
by the Sivers effect~\cite{Sive91}, i.e. a final state interaction between the
spectator system and the current quark jet. The data reported by the 
E704 $p^{\uparrow}p \rightarrow \pi X$ experiment show a similar
ambiguity. Moreover, as was stressed by Jaffe~\cite{Jaffe} the size
of the asymmetries observed by HERMES and E704 is somewhat large for
the (QCD based) expectation of a twist-3 effect. Makins~\cite{Makins} 
pointed out that future measurements on a transversely polarized target 
will be able to distinguish between the Sivers and Collins effects, as 
they give rise to a transverse single-spin asymmetry $A_{UT}$ of opposite 
sign for either process.

Further complications in this emerging field of transversity studies
may come from the Sudakov suppression of the Collins effect~\cite{Boer}.
However, this effect is expected to be small at the relatively low
$Q^2$ values exploited by the HERMES experiment. Moreover, Boer~\cite{Boer}
showed that the introduction of weighting factors (proportional to
$|p_{\perp}^{\pi}|/m_{\pi}$ with $p_{\perp}^{\pi}$ the transverse momentum
and $m_{\pi}$ the mass of the produced pion) could remove this sensitivity 
to the Sudakov suppression.

The promise of future transversity studies was probably best illustrated 
by two projects for the 'aggressive theorist' as proposed by Jaffe~\cite{Jaffe}:
(i) whether a relation exists between the difference of the transverse and
longitudinal spin distribution functions and the orbital angular momentum; and
(ii) whether the tensor charges (which can be derived from the transversity
distribution) is somehow related to chiral symmetry breaking. These projects
and the lively discussions on transversity during the workshop illustrate
the growing importance of this field. However, what is needed first are
high quality data and systematic comparisons of data collected in
different experiments in order to establish that the canonical framework
of hard QCD can be used when exploiting the new concept of transversity.
 
\subsection{Generalized parton distributions}

Generalized parton distributions (GPDs) are universal non-perturbative 
objects entering the description of hard exclusive electroproduction processes,
such as $e + p \rightarrow e + p + \gamma, \rho, \omega, \pi$, etc.
In the last few years, significant theoretical advances were made that
allow to factorize the amplitude for such processes into a hard scattering 
diagram at the parton level, on the one hand, and the GPDs on the other.
The internal quark-gluon structure
of hadrons is encoded in these distributions.

There are four different GPDs: $H$, $E$, $\tilde H$ and $\tilde E$.
Each of them depends on the same three variables: $x$, the average of the
longitudinal momentum fractions of the struck parton in its initial and final 
state, a skewdness parameter $\xi$ which measures the difference between 
these two momentum fractions, and $t$ the momentum transfer to the target 
nucleon.

Generalized parton distributions (GPDs) encompass both
the well-known parton distribution functions (PDFs) and the nucleon
form factors as limiting cases.
While the PDFs describe the probability to find a parton with fractional
momentum $x$ in the nucleon (forward scattering), the GPDs describe the
interference or correlation between two quarks with momentum fractions
$x+\xi$ and $x-\xi$ in the nucleon (off-forward scattering). Hence, the
GPDs are sensitive to partonic correlations.

The subject of GPDs was nicely introduced at the workshop in the opening
talk of Hoyer~\cite{Hoye02}, and further explained by Radyushkin~\cite{Rady02},
Mueller~\cite{Mull02} and Burkhardt~\cite{Burk02}. A number of interesting
observations were made, of which I mention some of the most important ones:
\begin{itemize}
\item{GPDs unify existing ways of describing hadronic structure, and
allow for accessing new information on partonic correlations.}
\item{GPDs allow the simultaneous determination of the longitudinal
momentum and transverse position of partons {\sl ("hadron tomography")}.}
\item{GPDs are sensitive to chiral symmetry breaking effects.}
\item{Educated guesses of GPDs result in predictions of cross sections
and single-spin asymmetries that are surprisingly close to the data.}
\end{itemize}
One can summarize these observations by noting that the GPDs present
a new powerful interface between fundamental calculations of hadronic
structure (such as lattice QCD and chiral-soliton models) and many 
different kinds of observables.

Many of these ideas were translated into practical calculations by
Vanderhaeghen~\cite{Vand02}, who stressed the importance of
measurements of GPDs in the domain $-\xi < x < \xi$, where the
sensitivity to $q\bar{q}$-correlations is particularly strong. 
As an example he showed a calculation of the beam charge asymmetry
$A_{ch}$ for deeply virtual Compton scattering (DVCS), where the size and
shape of the $\cos(\phi)$ dependence of $A_{ch}$ is entirely driven by these
$q\bar{q}$-correlations, which are usually referred to as the 'D-term'.
Ellinghaus~\cite{Elli02} presented first low-statistics data on $A_{ch}$ 
obtained at HERMES that are well described by these calculations.

The experimental investigation of DVCS represents just one avenue in
obtaining information on the GPDs. Guidal~\cite{Guid02} described how
systematic measurements of cross sections and asymmetries in DVCS, and
longitudinal meson electroproduction will enable us to obtain separate
data on the four known GPDs. He also introduced two new classes of
deeply virtual Compton scattering experiments:
\begin{itemize}
\item{Double DVCS (DDVCS) in which a virtual instead of a real photon
is produced in the final state. This process would make it possible to
study the $x$ and $\xi$ dependence of the GPDs separately, but has the
disadvantage that very high luminosities are required.}
\item{DVCS with a $\Delta$ resonance in the final state ($\Delta$VCS),
which can be identified by observing the $\Delta$ decay products
simultaneously with the produced photon. This process would make it
possible to estimate the $\Delta$ contribution to DVCS, which is
especially useful for experiments lacking sufficient resolution to
separate the $\Delta$ from the nucleon ground state.}
\end{itemize}

Examples of first measurements that
revealed the potential of the GPD framework were presented by
Hasch~\cite{Hasc02}, Elouadriri~\cite{Elou02} and Favart~\cite{Fava02}.
Although the data are still of modest quality, it is striking that
all cross sections and asymmetries measured (by ZEUS, H1, HERMES and
CLAS) are well reproduced by the available calculations. However, it
will require enormous improvements in experimental capabilities (both
in terms of energy resolution, and in particular luminosity) before it
will be possible to determine the $x$, $\xi$ and $t$ dependence of the
four GPDS from the data. Moreover, as the observables correspond to
integrals over (a sum of) GPDs, the extraction of $H$, $E$, $\tilde H$ 
and $\tilde E$ will always involve some model dependence. 
Korotkov~\cite{Koro02}, however, showed that under reasonable
assumptions the remaining model dependence is weak if $H(\xi,\xi,t)$ 
is extracted from DVCS measurements of single beam-spin asymmetries 
for small $t$.

Apart from their intrinsic interest (as probes of $q\bar{q}$-correlations), 
the determination of GPDs is needed to map out the angular momentum
structure of the nucleon. In 1997 Ji~\cite{Ji97} has shown that the first 
moment of $H$ and $E$ equals twice the total angular momentum
carried by the quarks and gluons -- in the limit $t \rightarrow$ 0.
By comparing the total angular momentum to the spin content of the
nucleon as measured in polarized deep-inelastic scattering experiments,
one get access to the (gauge dependent) orbital angular momentum of the
partons.

This finding triggered the enormous interest in the experimental
study of DVCS described above. I refer to the talks quoted above for
a more detailed account of these new data~\cite{Elli02,Hasc02,Elou02,Fava02}.
However, it was pointed out by Vanderhaeghen~\cite{Vand02} that transverse
spin asymmetries observed in longitudinal vector meson production also
reveal a surprising sensitivity to the total angular momentum carried
by the quarks. Within their chiral soliton model calculations, they showed
changes of the asymmetry from -0.15 to -0.30 if $J_u$ was changed from
0.1 to 0.4. It remains to be seen to what extend such measurements are
feasible in view of the required identification of the longitudinal
component of the reaction, $\gamma_L^* + p \rightarrow \rho_L^0 + p$,
and in view of possible model dependences.
 
\subsection{QCD effects in nuclear matter}

In her presentation on diffraction in $ep$-scattering, Abramovich~\cite{Abra02}
referred to two aspects of vector meson production which can be seen as
a link between the subjects discussed in the previous and present subsections:
\begin{itemize}
\item{Calculations for new ZEUS and H1 data on diffractive $\Upsilon$ 
production at $W \approx$ 140 GeV are shown to be highly sensitive to
the chosen GPD parameterization (up to a factor 2). Unfortunately, the data
are not yet sufficiently precise to distinguish between the various
calculations, but the example shows the large range of applicability
of the GPD framework.}
\item{When describing diffraction in terms of the interaction 
between a $q\bar{q}$-pair (originating from the hadronic structure of 
the virtual photon) and the target, it is assumed that the $q\bar{q}$-pair
has a small transverse size. As a result of its small size
the $q\bar{q}$-pair is assumed to represent a (white) color dipole, which
interacts only weakly with the proton. In other words Colour Transparency
is assumed to be valid.}
\end{itemize}
While Abramovich assumed the validity of Colour Transparency (CT), this striking
QCD prediction still has not been definitively demonstrated to exist
by experimental data. This became evident in the talk of
Strikman~\cite{Stri02}, who reviewed the present experimental evidence 
supporting this QCD prediction: (i) the observed A-dependence of 2-jet
production in pion induced experiments (E-791) at $E_{\pi}$ = 500 GeV
($A^{1.61\pm0.08}$) is in agreement with the CT-based prediction 
($A^{1.54}$); and (ii) the slope of the $t$-dependence of vector meson 
production shows the expected reduction with $Q^2$, albeit with poor 
statistics. At the workshop Borissov~\cite{Bori02} presented new additional
evidence in favour of colour transparency based on the analysis of coherent
$\rho^0$ production on $^{14}$N collected at HERMES.
By studying the $Q^2$-dependence of
the nuclear transparency in $\rho^0$ production while keeping the
coherence length constant, he found an increase with $Q^2$ of the
transparency by $0.081\pm0.027$ GeV$^{-2}$, which is consistent with 
the prediction based on colour-transparency of 0.07 GeV$^{-2}$. It is 
concluded that the experimental
evidence for this QCD prediction is finally accumulating, after many
false attempts in the past.

Another unverified QCD prediction discussed by Strikman~\cite{Stri02}
concerns the energy loss of partons propagating through nuclear matter.
According to QCD the energy loss per unit distance $dE/dL$ is proportional
to the distance traversed, which is different from our intuition based
on the Bethe-Bloch expression. In a recent paper Wang~\cite{Wang02}
used such a QCD approach to calculate the energy loss in hot and
cold nuclear matter. These results were discussed by Muccifora~\cite{Mucc02},
who presented semi-inclusive deep-inelastic scattering data collected by
the HERMES collaboration on various unpolarized nuclear targets. By comparing
the hadron yield per DIS event on nuclei to the same yield on deuterium,
a significant (energy dependent) attenuation is observed. This hadron
attenuation in $^{14}$N and $^{84}$Kr is well described by the aforementioned 
calculations of Wang if a value of $dE/dL \approx$ 0.3 GeV/fm is taken.
This value 
for the partonic energy loss in cold nuclear matter can be compared to
the energy loss derived from recent PHENIX data~\cite{PHEN01}
on $\pi^0$ production in Au+Au collisions at $\sqrt{s}$ = 130 GeV, 
yielding 0.25 GeV/fm~\cite{Wang02}. If the PHENIX
number is converted to the corresponding energy loss in the initial
hot stage of the Au+Au collision, a value of about 5 GeV/fm is found. 
Comparing this number to the value derived from the HERMES data for
cold nuclear matter, it was
concluded by Muccifora~\cite{Mucc02} that the gluon density (which drives
the energy loss) is a factor 15 higher in the initial phase of the Au+Au
collision. This result reflects a new synergy between two fields that used
to be essentially independent: relativistic heavy-ion collisions, and deep
inelastic scattering.

\subsection{Other recent developments}

Many other subjects were discussed at the workshop, not all of which can
be covered in this summary. On the experimental side, important reports on the
status of the COMPASS experiment~\cite{Lama02} at CERN and the RHIC
experiments~\cite{Bunc02} at BNL were presented. The COMPASS experiment
has carried out a successful commissioning run in late 2001, and real
data taking is expected to start in 2002. At BNL an important break-through
was reported, since it has been shown to be possible to inject and maintain
two polarized proton beams in RHIC at $\sqrt{s}$ = 100 GeV. These
technical developments imply that many more high-quality data are to
be expected in the nearby future.

On the theoretical side, very important developments in the field of lattice
gauge calculations were reported. Negele~\cite{Nege02} and 
Rakow~\cite{Rako02} described these results, which I summarize below:
\begin{itemize}
\item{The improved computing capabilities enable lattice QCD
calculations at smaller lattice spacing ($a$), large lattice
volumes ($L^3$) and -most importantly- smaller pion masses ($m_{\pi}$).}
\item{The differences between quenched (no sea quarks) and unquenched
calculations turn out to be minimal.}
\item{In order to make much more realistic extrapolations to light pion 
(or quark) masses Chiral Perturbation Theory is successfully invoked.}
\end{itemize}
Although many improvements are still needed, it is gratifying to
see that lattice QCD now is able to reproduce many experimental
observables such as moments of parton distributions.

\section{FUTURE PERSPECTIVES}

The developments described in the previous section imply that the
study of the QCD structure of the nucleon is entering a new phase.
Many new high-precision data will become available in the nearby 
future from HERMES, JLab, COMPASS and RHIC-spin. At the same
time the framework of generalized parton distributions has
extended dramatically the number of well-defined observables
that can be used to compare experimental results with theoretical
calculations such as those based on the chiral-soliton (or instanton)
models or advanced lattice QCD calculations. This atmosphere of 
progress and anticipation was clearly present at the workshop.

More specifically the following experimental results can be
expected in the next 5 years (or so):
\begin{itemize}
\item{{\sl Flavour decomposition of nucleon spin:} Precise spin-dependent
distribution functions for separate flavours, i.e. $\Delta u(x)$,
$\Delta d(x)$, $\Delta \bar{u}(x)$, $\Delta \bar{d}(x)$, and
$\Delta s(x)$ will be produced by HERMES, COMPASS and RHIC spin.}
\item{{\sl Gluon Polarization:} following the pioneering measurements
at HERMES~\cite{Herm00} considerably more precise data on $\Delta G/G$
are expected from COMPASS, SLAC experiment E161 and RHIC spin.}
\item{{\sl Transversity:} measurements with transversely polarized
targets are scheduled at HERMES in the next couple of years, and
will soon be followed by similar measurements at COMPASS and (somewhat
later) at RHIC. Hence, first (but by no means complete) measurements
of the $u$-quark transversity distribution $\delta u(x)$ will be
available.}
\item{{\sl Generalized Parton Distributions:} measurements of
deeply virtual Compton scattering and exclusive (vector) meson
production will be continued at JLab and HERMES. Many reaction
channels will be explored, but the kinematic ranges and luminosities
available will severely limit the extraction of the complete
$x$, $\xi$ and $t$ dependence of the GPDs.}
\item{{\sl Search for missing resonances and hybrids:} although
not covered at the workshop, the search for new baryon
resonances and hybrids also constitutes an important avenue
in studying the QCD structure of nucleons and baryons. At present,
and in the nearby future, these searches are being conducted at JLab 
(CLAS), ELSA, GRAAL and SPRING8, for instance. These facilities
(soon supplemented by MAMI-C) will provide good data in the
non-charm sector.}
\end{itemize}
 
Hence, a wealth of data is to be expected in the coming years.
If these data are accompanied by similar theoretical efforts,
considerable progress can already be expected at a time scale
of 5 to 7 years.

\newpage

\section{A ROAD MAP}

Although many results are expected in the next 5 years, it is also
clear today that many questions will remain unanswered. As an example
the study of generalized parton distributions can be mentioned,
which will require new electron-scattering facilities of high
luminosity. Similarly, the determination of the tensor charge
of the nucleon through the measurement of the transversity 
distribution will also require high precision measurements,
which cannot be carried out at existing facilities.

With these arguments in mind, the end of the workshop was
devoted to presentations and a subsequent discussion on 
future facilities. Rather than
summarizing the physics objectives of each of these facilities,
which one can find in the preceding papers, I decided to focus 
on answering the key question for which the workshop was
organized, i.e. where and how to go in the next decade? It was
the aim of this workshop to arrive at a {\sl Road Map} of the
type that can be found in OECD reports describing future
plans in high-energy physics, for instance~\cite{OECD02}.

Following the example of Ref.~\cite{OECD02} I sketched the following
road map, admitting that some of the choices carry a personal bias:
\begin{enumerate}
\item{Exploit current frontier facilities in our field such as
COMPASS, HERMES-II, JLab and RHIC-spin, until surpassed by new
facilities.}
\item{Form a coherent long-range plan shared by the entire
community in our field. This is of particular importance in
Europe, where no such tradition exists.}
\item{Prepare for the approval and construction ({\sl before 2008}):
	\begin{itemize}
	\item{The JLab 12 GeV upgrade~\cite{Meck02} including the equipment for
	the proposed new real photon experiments (Hall D)\footnote{
		In this respect one could also mention the European
		high-energy storage ring (HESR) project at GSI~\cite{HESR02},
		which will provide $1.5 \div 15$ GeV/c (cooled) anti-proton 
		beams impinging on fixed (internal) targets. This project
		was not discussed at the workshop, but has a similar
		scope as the JLab 12 GeV project. }.
	     }
	\end{itemize}	
     }
\item{Establish a vigorous R\&D program to fully develop ({\sl for exploitation
after 2008}):
	\begin{itemize}
	\item{A fixed target $eN$ machine in Europe with polarized beams and
	targets~\cite{VonH02}, $E_e = 25 \div 100$ GeV, and a luminosity 
	of about 10$^{35}$ N/cm$^2$/s.}
	\item{The electron-ion collider (EIC) at BNL~\cite{Miln02} with 
	both beams polarized, $\sqrt{s} \approx 30 \div 100$ GeV, and a 
	luminosity of about
	10$^{33}$ N/cm$^2$/s.}
	\end{itemize}
     }
\end{enumerate}

The project mentioned for the mid-term future is an existing proposal, 
which has already been given to the funding agencies. The required 
investments are modest as compared to the budgets required for the 
projects listed under item 4, and the physics objectives concern the 
search for hybrids and missing resonances\footnote{
	Actually, similar arguments apply to the HESR facility. However,
	while the JLab project concentrates on the $uds$-quark sector,
	the HESR project focusses on the charm sector.}.

The most important and most challenging projects for studying the QCD 
structure of the nucleon are listed under item 4. Also in this case there
is a nice a complementarity between the EU- and US-plans: while EIC exploits
the benefits of a collider experiment, the European plans have the
advantage of a higher luminosity. Several scenarios already exist for a
high-intensity $eN$-facility in Europe~\cite{TESL01}. Moreover, at the 
workshop, a novel
scenario was discussed by von Harrach~\cite{VonH02}, in which use was 
made of the existing infrastructure of the HERA ring at DESY. Although 
the accelerator 
configurations proposed in the various scenarios are different, the physics 
motivation is the same; the study of semi-inclusive reactions to determine the
transversity distributions, and the study of exclusive reactions
to access the Generalized Parton Distributions, requiring in both
cases high-intensity (polarized) electron beams and polarized targets.  
For the semi-inclusive studies the optimal beam energy range is 
$50 \div 100$ GeV, while beam energies of $25 \div 50$ GeV are more 
suitable for extracting cross sections and their scale dependence in
exclusive measurements. The new European initiative~\cite{Ferr02}
discussed at the workshop aims at incorporating both objectives
into one new proposal.

In the nearby future, item 2 of the {\sl Road Map} given above is probably
the most important one. Within Europe a unified view needs to be developed
regarding future QCD studies of the nucleon. In this respect it is very 
fortunate that NuPECC has initiated the development of a new long range 
plan with one working group dedicated to the study of QCD~\cite{Weis02}. 
As input to this working group it is of great significance that
a large fraction of our community has expressed its
support for the development of a high-intensity $eN$-facility in
Europe by signing the {\sl Declaration of Ferrara}~\cite{Ferr02}.
For this initiative and the excellent organization of the workshop
QCD-N'02 the organizers, and more in particular Wolf-Dieter Nowak
and Enzo de Sanctis, deserve all praise.

\end{document}